\documentclass[english,twocolumn,showpacs,preprintnumbers,amsmath,amssymb]{revtex4}
\usepackage[T1]{fontenc}
\usepackage[latin1]{inputenc}
\usepackage{graphicx}

\makeatletter

\newcommand{\noun}[1]{\textsc{#1}}

\makeatletter

\usepackage{dcolumn}
\usepackage{bm}

\makeatother

\usepackage{babel}
\makeatother
\begin{document}

\title{Relaxation dynamics in strained fiber bundles}

\author{Srutarshi Pradhan}

\email{pradhan.srutarshi@ntnu.no}

\author{Per C. Hemmer}

\email{per.hemmer@ntnu.no}

\affiliation{Department of Physics, Norwegian University of Science and Technology,
N--7491 Trondheim, Norway}

\begin{abstract}
Under an applied external load the global load-sharing fiber bundle
model, with individual fiber strength thresholds sampled randomly
from a probability distribution, will relax to an equilibrium state,
or to complete bundle breakdown. The relaxation can be viewed as taking
place in a sequence of steps. In the first step all fibers weaker
than the applied stress fail. As the total load is redistributed on
the surviving fibers, a group of secondary fiber failures occur, etc.
For a bundle with a finite number of fibers the process stops after
a finite number of steps, $t$. By simulation and theoretical estimates,
it is determined how $t$ depends upon the stress, the initial load
per fiber, both for subcritical and supercritical stress. The two-sided
critical divergence is characterized by an exponent $-1/2$, independent
of the probability distribution of the fiber thresholds.\\

\end{abstract}


\maketitle

\section{Introduction}

Bundles of fibers, with a statistical distribution of breakdown thresholds
for the individual fibers, are simple and interesting models of failure
processes in materials. They can be analyzed to an extent that is
not possible for most materials (for reviews, see \cite{Herrmann,Chakrabarti,Sahimi,Sornette,Bhattacharyya}).

We consider a bundle with a large number $N$ of elastic and parallel
fibers, clamped at both ends. When the load on fiber $i$ is increased
beyond a threshold value $x_{i}$, the fiber ruptures. The breakdown
thresholds $x_{i}$ for the separate fibers are assumed to be independent
random variables with a probability density $p(x)$, and a corresponding
cumulative distribution function $P(x)$: \begin{equation}
\mbox{ Prob }(x_{i}\leq x)\equiv P(x)=\int_{0}^{x}\; p(y)\; dy.\end{equation}
 The mechanism for how the extra stress caused by a fiber failure
is redistributed among the unbroken fibers must be specified. We study
here the classical version, the equal-load-sharing model, in which
a ruptured fiber carries no load, and the increased stress caused
by a failed element is shared equally by all the remaining intact
fibers in the bundle \cite{Peirce}.

If an external load $F$ is applied to a fiber bundle, the resulting
failure events can be seen as a sequential process \cite{PC,PBC,BPC}.
In the first step all fibers that cannot withstand the applied load
break. Then the stress is redistributed on the surviving fibers, which
compels further fibers to fail, etc. This iterative process continues
until all fibers fail, or an equilibrium situation with a nonzero
bundle strength is reached. Since the number of fibers is finite,
the number of steps, $t$, in this sequential process is {\em finite}.
In this paper we determine how $t$ depends upon the number of fibers
and, more importantly, upon the stress $\sigma$, the applied external
load per fiber, \begin{equation}
\sigma=F/N.\end{equation}

At a force $x$ per surviving fiber, the total force on the bundle
is $x$ times the number of intact fibers. The expected or average
force at this stage is therefore \begin{equation}
F(x)=N\, x\,(1-P(x)).\label{load}\end{equation}
One may consider $x$ to represent the elongation of the bundle, with
the elasticity constant set equal to unity. The maximum $F_{c}$ of
$F(x)$ corresponds to the value $x_{c}$ for which $dF/dx$ vanishes.
Thus \begin{equation}
1-P(x_{c})-x_{c}p(x_{c})=0.\end{equation}
 We characterize the state of the bundle as \textit{subcritical} or
\textit{supercritical} depending upon the stress value relative to
the critical stress \begin{equation}
\sigma_{c}=F_{c}/N,\end{equation}
above which the bundle collapses completely. Critical properties of
fiber bundles have been discussed before, but with a signature different
from the one that we use here, and always with the critical point
approached from the subcritical side \cite{Sornette,Hansen,PBC}.
The function $t(\sigma)$ that we focus on, however, exhibits critical
divergence when the critical point is approached from \textit{either}
side. As an example, we show in Fig.\ 1 $t(\sigma)$ obtained by
simulation for a uniform threshold distribution.

\begin{center}\includegraphics[width=3in,height=3in]{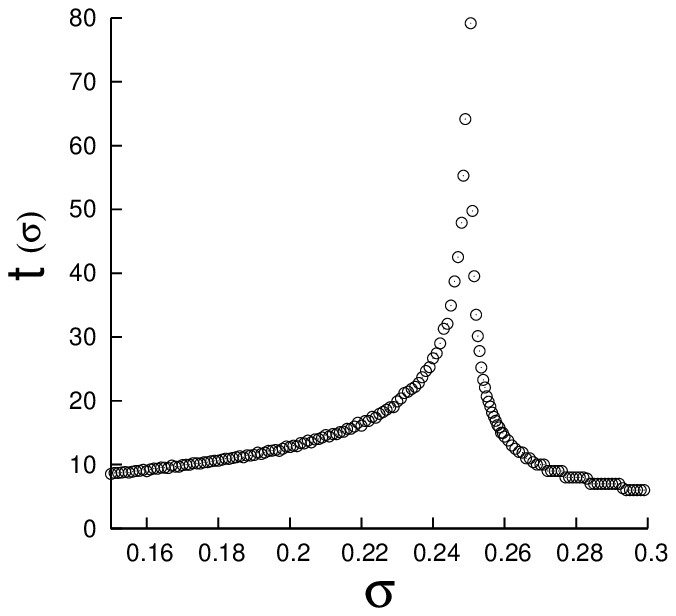}\par\end{center}

{\footnotesize FIG.\ 1. Number of relaxation steps $t(\sigma)$ for
a fiber bundle with a uniform threshold distribution (\ref{uniform}).
Here $\sigma_{c}=0.25$. The figure is based on 1000 samples, each
with $N=10^{6}$ fibers.}\\

We study the stepwise failure process in the bundle, when a fixed
external load $F=N\sigma$ is applied. Let $N_{t}$ be the number
of intact fibers at step no.\ $t$, with $N_{0}=N$. We want to determine
how $N_{t}$ decreases until the degradation process stops. With $N_{t}$
intact fibers, an expected number \begin{equation}
\left[NP(N\sigma/N_{t})\right]\end{equation}
 of fibers will have thresholds that cannot withstand the load, and
consequently these fibers break immediately. Here $[X]$ denotes the
largest integer not exceeding $X$. The number of intact fibers in
the next step is therefore \begin{equation}
N_{t+1}=N-\left[NP(N\sigma/N_{t})\right].\label{Nt}\end{equation}
 Since $N$ is a large number, the ratio \begin{equation}
n_{t}=\frac{N_{t}}{N}\end{equation}
 can for most purposes be considered a continuous variable. By (\ref{Nt})
we have essentially \cite{PC,PBC,BPC} \begin{equation}
n_{t+1}=1-P(\sigma/n_{t}).\label{n}\end{equation}

In Sec.\ II we study $t(\sigma)$ in the supercritical domain, while
Sec.\ III is devoted to subcritical situations. Simulation results
are presented for two threshold distributions, the uniform and a Weibull
distribution, and these are compared with detailed analytic results.
The theoretical analysis is, however, not limited to these special
threshold distributions. In Sec.\ IV we summarize our results and
discuss briefly the approximations involved.

\section{Supercritical relaxation}

We investigate first the supercritical situation, $\sigma>\sigma_{c}$,
with positive values of \begin{equation}
\epsilon=\sigma-\sigma_{c},\end{equation}
 and start with the simplest model.

\subsection{Uniform threshold distribution}

Consider a bundle in which the failure thresholds are distributed
according to the uniform distribution \begin{equation}
P(x)=\left\{ \begin{array}{cl}
x & \mbox{ for }0\leq x\leq1\\
0 & \mbox{ for }x>1\end{array}\right.\label{uniform}\end{equation}
 For this case the load curve (\ref{load}) is parabolic, \begin{equation}
F=Nx(1-x),\end{equation}
 with the critical point at $x_{c}=1/2$, $\sigma_{c}=1/4$. Simulation
results for the uniform threshold distribution are presented in Fig.\ 1.

The basic equation (\ref{n}) takes the form \begin{equation}
n_{t+1}=1-\frac{\sigma}{n_{t}}=1-\frac{\frac{1}{4}+\epsilon}{n_{t}}.\label{ntuni}\end{equation}
 This nonlinear iteration can be transformed into a linear one by
the following procedure. Introduce first \begin{equation}
n_{t}={\textstyle \frac{1}{2}}-y_{t}\sqrt{\epsilon},\end{equation}
 into (\ref{ntuni}), with a result that may be written \begin{equation}
\frac{y_{t+1}-y_{t}}{1+y_{t}y_{t+1}}=2\sqrt{\epsilon}.\end{equation}
 Putting \begin{equation}
y_{t}=\tan v_{t},\end{equation}
 we have \begin{equation}
2\sqrt{\epsilon}=\frac{\tan v_{t+1}-\tan v_{t}}{1+\tan v_{t+1}\;\tan v_{t}}=\tan(v_{t+1}-v_{t}).\end{equation}
 Hence $v_{t+1}-v_{t}=\tan^{-1}(2\sqrt{\epsilon})$, with solution
\begin{equation}
v_{t}=v_{0}+t\;\tan^{-1}(2\sqrt{\epsilon}).\end{equation}
 In the original variable the solution reads {\small \begin{eqnarray}
n_{t} & = & {\textstyle \frac{1}{2}}-\sqrt{\epsilon}\;\tan\left(\tan^{-1}(\frac{\frac{1}{2}-n_{0}}{\sqrt{\epsilon}})+\; t\;\tan^{-1}(2\sqrt{\epsilon})\right)\\
 & = & {\textstyle \frac{1}{2}}-\sqrt{\epsilon}\;\tan\left(-\tan^{-1}(1/2\sqrt{\epsilon})+\; t\;\tan^{-1}(2\sqrt{\epsilon})\right),\label{finaluni}\end{eqnarray}
}where $n_{0}=1$ has been used.

Eq.\ (\ref{ntuni}) shows that when $n_{t}$ obtains a value in the
interval $(0,\sigma)$, the next iteration gives complete bundle failure.
Taking $n_{t}=\sigma$ as the penultimate value gives a lower bound,
$t_{l}$, for the number of iterations, while using $n_{t}=0$ in
(\ref{finaluni}) gives an upper bound $t_{u}$. Adding unity for
the final iteration, (\ref{finaluni}) gives the bounds \begin{equation}
t_{u}(\sigma)=1+\frac{2\tan^{-1}(1/2\sqrt{\epsilon})}{\tan^{-1}(2\sqrt{\epsilon})},\label{tu}\end{equation}
 and \begin{equation}
t_{l}(\sigma)=1+\frac{\tan^{-1}((\frac{1}{4}-\epsilon)/\sqrt{\epsilon})+\tan^{-1}(1/2\sqrt{\epsilon})}{\tan^{-1}(2\sqrt{\epsilon})}.\label{tl}\end{equation}
 It is easy to show that $t_{u}(\sigma)-t_{l}(\sigma)=1$. In Fig.\ 2A
we show that these bounds nicely embrace the simulation results.

\begin{center}\includegraphics[width=3in,height=2.5in]{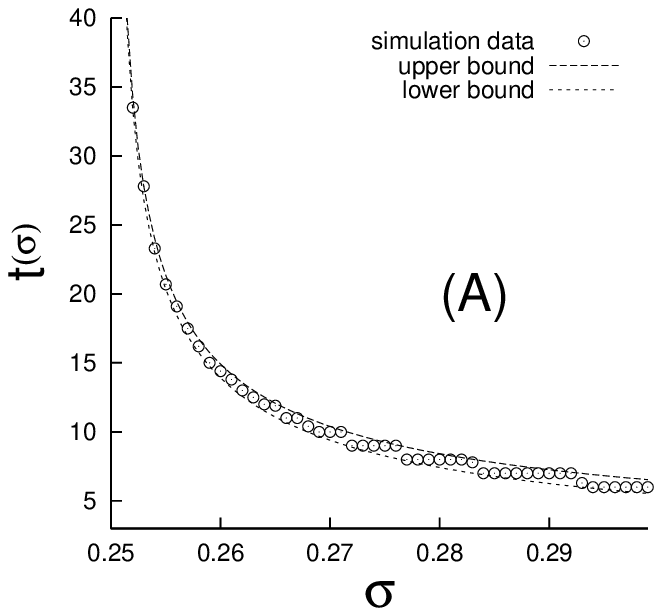}\par\end{center}

\begin{center}\includegraphics[width=3in,height=2.5in]{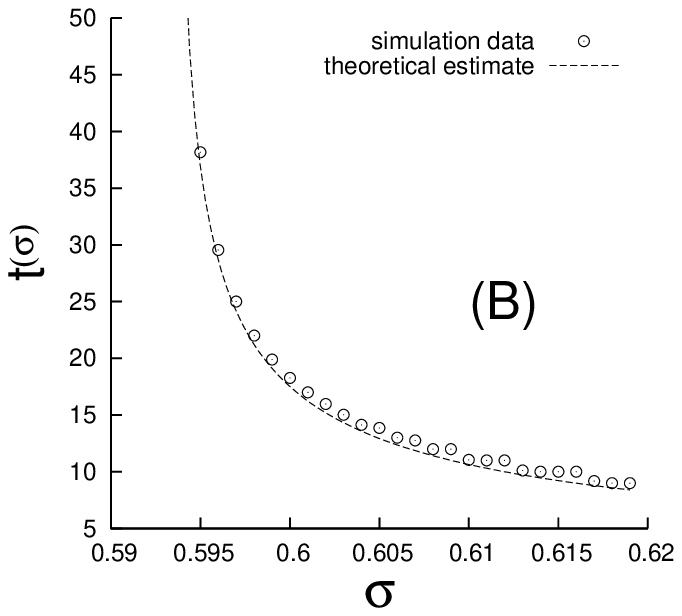}\par\end{center}

{\footnotesize FIG.\ 2. Simulation results with supercritical stress
for (A) the uniform threshold distribution (\ref{uniform}), and (B)
the Weibull distribution (\ref{Weibull}). The graphs are based on
10000 samples with $N=10^{6}$ fibers in each bundle. The dashed lines
represent the theoretical estimates (\ref{tu}), (\ref{tl}) and (\ref{estimateW})}.
\\

Note that both the upper and the lower bound behave as $\epsilon^{-\frac{1}{2}}$
for small $\epsilon$. A rough approximation near the critical point
is \begin{equation}
t(\sigma)\approx{\textstyle \frac{1}{2}}\pi(\sigma-\sigma_{c})^{-\frac{1}{2}}.\end{equation}

\subsection{General threshold distributions}

The uniform distribution is amenable to analysis to a degree not shared
by other threshold distributions. Therefore we now discuss how to
handle other distributions, and we start by a special case, a Weibull
distribution of index $5$, \begin{equation}
P(x)=1-e^{-x^{5}},\hspace{1cm}x\geq0.\label{Weibull}\end{equation}
 The critical parameters for this case are $x_{c}=5^{-1/5}=0.72478$
and $\sigma_{c}=(5e)^{-1/5}=0.5933994$. Simulation results for $t(\sigma)$
are displayed in Fig.\ 2B for the Weibull supercritical case. The
variation with the external stress $\sigma$ is qualitatively similar
to the results for the uniform threshold distribution.

The interesting values of the external stress are close to $\sigma_{c}$,
because for large supercritical stresses the bundle breaks down almost
immediately. For $\sigma$ slightly above $\sigma_{c}$ the iteration
function \begin{equation}
n_{t+1}=f(n_{t})=1-P(\sigma/n_{t})=e^{-(\sigma/n_{t})^{5}},\label{it-W}\end{equation}
 takes the form sketched in Fig.\ 3.

\begin{center}\includegraphics[width=3in,height=2.5in]{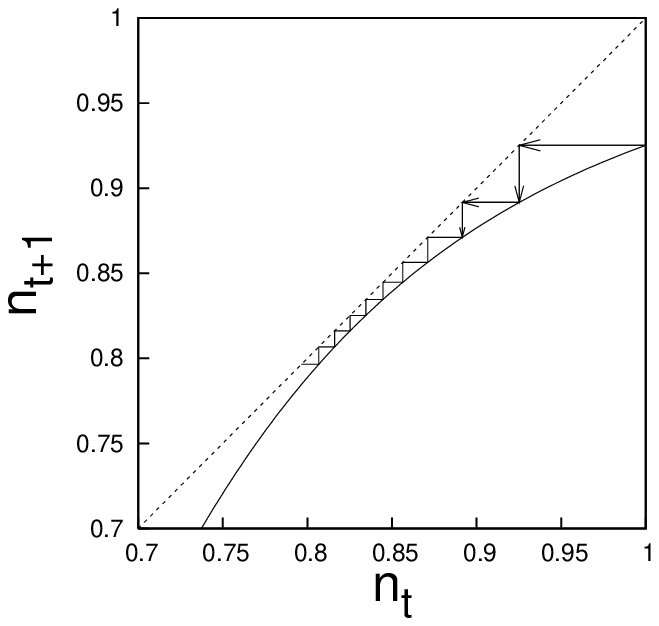}\par\end{center}

{\footnotesize FIG.\ 3. The iteration function $f(n)$ for the Weibull
distribution (\ref{Weibull}). Here $\sigma=0.6$, slightly greater
than the critical value $\sigma_{c}=0.5933994.$}\\

The iteration function is almost tangent to the reflection line $n_{t+1}=n_{t}$
and a long channel of width proportional to $\epsilon$ appears. The
dominating number of iterations occur within this channel (see Fig.
3). The channel wall formed by the iteration function is almost parabolic
and is well approximated by a second-order expression \begin{equation}
n_{t+1}=n_{c}+(n_{t}-n_{c})+a(n_{t}-n_{c})^{2}+b(\sigma_{c}-\sigma).\label{quadr}\end{equation}
 Here $n_{c}=e^{-1/5}$ is the fixed point, $n_{t+1}=n_{t}$, of the
iteration at $\sigma=\sigma_{c}$. With $u=(n-n_{c})/b$ and $\epsilon=\sigma-\sigma_{c}$
(\ref{quadr}) takes the form \begin{equation}
u_{t+1}-u_{t}=-Au_{t}^{2}-\epsilon,\end{equation}
 with $A=ab$. In the channel $u$ changes very slowly, so we may
treat the difference equation as a a differential equation: \begin{equation}
\frac{du}{dt}=-Au^{2}-\epsilon,\end{equation}
 with solution \begin{equation}
t\sqrt{A\epsilon}=-\tan^{-1}\left(u\sqrt{A/\epsilon}\right)+\mbox{ constant }.\end{equation}
 Thus \begin{equation}
t_{f}-t_{i}=(A\epsilon)^{-\frac{1}{2}}\left\{ \tan^{-1}(u_{i}\sqrt{A/\epsilon})-\tan^{-1}(u_{f}\sqrt{A/\epsilon})\right\} \label{tf-ti}\end{equation}
 is the number of iterations \textit{in the channel}, starting with
$u_{i}$, ending with $u_{f}$. This treatment is general and can
be applied to any threshold distribution near criticality. Although
the vast majority of the iterations occur in the channel, there are
a few iterations at the entrance and at the exit of the channel that
may require attention in special cases. The situation is similar to
type I intermittency in dynamical systems\cite{Pomeau}, but in our
case the channel is traversed merely once.

For the Weibull distribution the expansion (\ref{quadr}) has the
precise form\begin{eqnarray}
n_{t} & = & e^{-(\sigma/n)^{5}}\simeq e^{-1/5}+(n-n_{c})\nonumber \\
 &  & -{\textstyle \frac{5}{2}}e^{1/5}(n-n_{c})^{2}-5^{1/5}(\sigma-\sigma_{c}),\label{b1}\end{eqnarray}
where $n_{c}=e^{-1/5}$, $a={\textstyle \frac{5}{2}}e^{1/5}$, $b=5^{1/5}$
and $A={\textstyle \frac{5}{2}}(5e)^{1/5}$. For completeness we must
also consider the number of iteration to reach the entrance to the
channel. It is not meaningful to use the quadratic approximation (\ref{b1})
where it is not monotonously increasing, i.e. for $n>n_{m}=n_{c}+1/(2a)=\frac{6}{5}e^{-1/5}\simeq0.98$.
Thus we take $n_{i}=n_{m}$ as the entrance to the channel, and add
one extra iteration to arrive from $n_{0}=1$ to the channel entrance.
(Numerical evidence for this extra step: For $\sigma=\sigma_{c}$
the iteration (\ref{it-W}) starts as follows: $n_{0}=1.00$, $n_{1}=0.93$,
$n_{2}=0.90$, while using the quadratic function with $n_{0}=n_{m}=0.98$
as the initial value, we get after one step $n_{1}=0.90$, approximately
the same value that the exact iteration reaches after two steps.)
With $n_{f}=0$ we obtain from (\ref{tf-ti}) in the Weibull case
the estimate\begin{eqnarray}
t & = & 1+(A\epsilon)^{-1/2}\left\{ \tan^{-1}(e^{-1/5}\sqrt{A/\epsilon}\,/5b)\right.\nonumber \\
 &  & \left.+\tan^{-1}(e^{-1/5}\sqrt{A/\epsilon}\,/b)\right\} ,\label{estimateW}\end{eqnarray}
with $A=\frac{5}{2}(5e)^{1/5}$ and $b=5^{1/5}$.

We see in Fig.\ 2B that the theoretical estimate (\ref{estimateW})
gives an excellent representation of the simulation data. Near the
critical point (\ref{estimateW}) has the asymptotic form \begin{equation}
t\approx\pi(A\epsilon)^{-1/2}\propto(\sigma-\sigma_{c})^{-1/2},\label{div}\end{equation}
 with the same critical index as for the uniform threshold distribution.
The divergence is caused by the large number of iterations in the
narrow channel in Fig.\ 3. For a general threshold distribution such
a channel will always be present, and therefore the divergence (\ref{div})
is universal. Moreover, the amplitude of $(\sigma-\sigma_{c})^{-1/2}$,
as well as an excellent representation of the complete $t(\sigma)$
function may, for a given threshold distribution, be obtained by the
same procedure as used above for the Weibull case.

\section{Subcritical relaxation}

We now assume the external stress to be subcritical, $\sigma<\sigma_{c}$,
and introduce the positive parameter \begin{equation}
\varepsilon=\sigma_{c}-\sigma\end{equation}
 to characterize the deviation from the critical point.

Also in the subcritical situation a bundle with the uniform threshold
distribution (\ref{uniform}) can be analyzed analytically to a greater
extent than for other distributions, and consequently we start with
this case.

\subsection{Uniform threshold distribution}

Using a similar method as in the supercritical situation we introduce
into (\ref{ntuni}) \begin{equation}
n_{t}={\textstyle \frac{1}{2}}+\sqrt{\varepsilon}/z_{t},\end{equation}
 as well as $\sigma=\frac{1}{4}-\varepsilon$, with the result \begin{equation}
2\sqrt{\varepsilon}=\frac{z_{t+1}-z_{t}}{1-z_{t+1}\; z_{t}}.\end{equation}
 In this case \begin{equation}
z_{t}=\tanh w_{t}\end{equation}
 is the useful substitution. It gives \begin{equation}
2\sqrt{\varepsilon}=\frac{\tanh w_{t+1}-\tanh w_{t}}{1-\tanh w_{t+1}\;\tanh w_{t}}=\tanh(w_{t+1}-w_{t}).\end{equation}
 Thus $w_{t+1}-w_{t}=\tanh^{-1}(2\sqrt{\varepsilon})$, i.e. \begin{equation}
w_{t}=w_{0}+t\;\tanh^{-1}(2\sqrt{\varepsilon}).\end{equation}
 Starting with $n_{0}=1$, we obtain $z_{0}=2\sqrt{\varepsilon}$
and hence \begin{equation}
w_{t}=(1+t)\;\tanh^{-1}(2\sqrt{\varepsilon}).\end{equation}
 This corresponds to \begin{equation}
n_{t}={\textstyle \frac{1}{2}}+\frac{\sqrt{\varepsilon}}{\tanh\left\{ (1+t)\tanh^{-1}(2\sqrt{\varepsilon})\right\} }\label{ntor}\end{equation}
 in the original variable.

Apparently $n_{t}$ reaches a fixed point $n^{*}=\frac{1}{2}+\sqrt{\varepsilon}$
after an infinite number of iterations. However, our bundle contains
a \textit{finite} number of fibers, and therefore only a finite number
of steps is needed for the iteration to arrive at a fixed point $N^{*}$
of the integer iteration (\ref{Nt}), \begin{equation}
N_{t+1}=N-[\sigma N^{2}/N_{t}].\label{intuni}\end{equation}
 Since $X\leq[X]<X+1$ a fixed point $N^{*}$ of (\ref{intuni}) must
satisfy \cite{PBC} {\small \begin{equation}
\frac{N}{2}\left(1+\sqrt{1-4\sigma}\right)\leq N^{*}<\frac{1}{2}\left(N+1+\sqrt{N^{2}(1-4\sigma)+2N+1}\right).\label{Nfix}\end{equation}
} It is interesting to note that (\ref{intuni}) has in general several
fixed points for a given value of $\sigma$. With $N=10^{6}$ and
$\sigma=0.249$, for instance, there are nine fixed points, viz.\ $531623,531624,\ldots,531631$,
the complete set of integers within the interval (\ref{Nfix}). Since
our iteration starts high at $N_{0}=N$, with steadily decreasing
values of $N_{t}$, it will stop at the \textit{upper} fixed point,
the largest integer satisfying (\ref{Nfix}).

As long as $N(1-4\sigma)\gg1$, which is fulfilled in our simulations,
we may take \begin{equation}
N_{{\rm u}}^{*}=\frac{N}{2}\left(1-\sqrt{1-4\sigma}\right)+\frac{1}{2}\left(1+(1-4\sigma)^{-1/2}\right)\label{app}\end{equation}
 as a good approximation to the upper fixed point (in the example
above (\ref{app}) gives $531631.1$, compared with $N_{{\rm u}}^{*}=531631$).

As a consequence we use \begin{equation}
n_{t}=\frac{N_{{\rm u}}^{*}}{N}=\frac{1}{2}+\sqrt{\varepsilon}+\frac{1}{4N}\left(2+\varepsilon^{-1/2}\right)\end{equation}
 as the final value in (\ref{ntor}). Consequently we obtain the following
estimate for the number of iterations to reach this value: \begin{equation}
t(\sigma)=-1+\frac{\coth^{-1}\left\{ 1+(1+2\sqrt{\varepsilon})/4N\varepsilon\right\} }{\tanh^{-1}(2\sqrt{\varepsilon})}.\label{subnorm}\end{equation}
 We see in Fig.\ 4A that the simulation data are well approximated
by the analytic formula (\ref{subnorm}).

For very large $N$ (\ref{subnorm}) is approximated by \begin{equation}
t=\frac{\ln(8N\varepsilon)}{4}\;\varepsilon^{-1/2}.\end{equation}
 The critical behavior is again characterized by a square root divergence,
in this case somewhat modified by a logarithmic term.

\subsection{General threshold distribution}

Again we use the Weibull distribution (\ref{Weibull}) as an example
threshold distribution. Simulation results for the subcritical Weibull
distribution are shown in Fig.\ 4B.

\begin{center}\includegraphics[width=3in,height=2.5in]{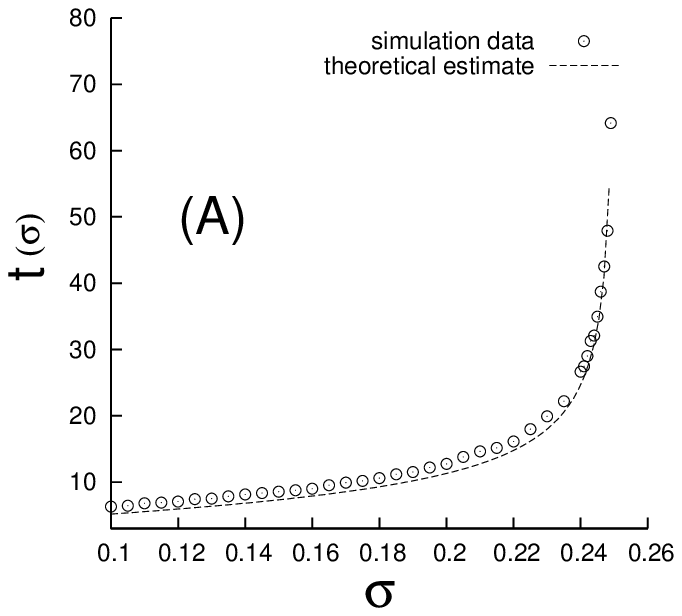}\par\end{center}

\begin{center}\includegraphics[width=3in,height=2.5in]{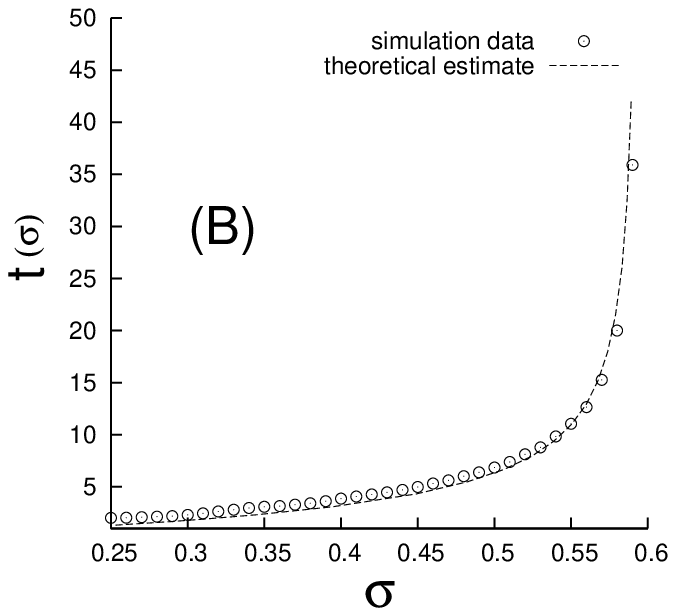}\par\end{center}

{\footnotesize FIG.\ 4. Simulation results with subcritical stress
for (A) the uniform threshold distribution, and (B) the Weibull distribution
(\ref{Weibull}). The graphs are based on $10000$ samples with $N=10^{6}$
fibers in each bundle. The dotted lines are the theoretical estimates,
eq.\ (\ref{subnorm}) for the uniform distribution case, and eqs.\ (\ref{par1}-\ref{par2})
for the Weibull case.} \\

Forgetting for the moment the finiteness of the fiber bundle, the
iteration (\ref{n}), \begin{equation}
n_{t+1}=1-P(\sigma/n_{t}),\end{equation}
 will reach a fixed point $n^{*}$ after infinite many steps. The
deviation from the fixed point, $n_{t}-n^{*}$, will decrease exponentially
near the fixed point \cite{PC,PBC,BPC,PAS}: \begin{equation}
n_{t}-n^{*}\propto e^{-t/\tau},\label{exp}\end{equation}
 with \begin{equation}
\tau=1/\ln\left\{ n^{*2}\sigma^{-1}/p(\sigma/n^{*})\right\} .\end{equation}

For the Weibull threshold distribution, in particular, \begin{equation}
p(\sigma/n^{*})=5(\sigma/n^{*})^{4}\;\exp\left(-(\sigma/n^{*})^{5}\right)=5\sigma^{4}/n^{*3},\end{equation}
 and thus \begin{equation}
\tau=1/\ln(n^{*5}/5\sigma^{5})\end{equation}
 for the Weibull case. If we allow ourselves to use the exponential
formula (\ref{exp}) all the way from $n_{0}=1$, we obtain \begin{equation}
n_{t}-n^{*}=(1-n^{*})e^{-t/\tau}.\label{fixW}\end{equation}

For a finite number $N$ of fibers the iteration will stop after a
finite number of steps. It is a reasonable supposition to assume that
the iteration stops when $N_{t}-N^{*}$ is of the order $1$. This
corresponds to take the left-hand side of (\ref{fixW}) equal to $1/N$.
The corresponding number of iterations is then given by \begin{equation}
t=\tau\;\ln\left(N(1-n^{*})\right)=\frac{\ln\left(N(1-n^{*})\right)}{\ln(n^{*5}/5\sigma^{5})}.\label{tWsub}\end{equation}
 Solving the Weibull iteration $n^{*}=\exp(-(\sigma/n^{*})^{5})$
with respect to $\sigma$ and inserting into (\ref{tWsub}), we obtain
\begin{eqnarray}
t & = & -\frac{\ln\left\{ N(1-n^{*})\right\} }{\ln\left\{ 5(-\ln n^{*})\right\} }\label{par1}\\
\sigma & = & n^{*}(-\ln n^{*})^{1/5}.\label{par2}\end{eqnarray}
 These two equations represent the function $t(\sigma)$ on parameter
form, with $n^{*}$ running from $n_{c}=e^{-1/5}$ to $n^{*}=1$.
In Fig.\ 4B this theoretical estimate is compared with the simulation
data. The agreement is satisfactory.

For $n^{*}=n_{c}=e^{-1/5}$ (\ref{par1}) shows that $t$ is infinite,
as it should be. To investigate the critical neighborhood we put $n^{*}=n_{c}(1+\xi)$
with $\xi$ small, to obtain to lowest order \begin{eqnarray}
t & = & {\cal O}(\xi^{-1})\\
\sigma_{c}-\sigma & = & {\cal O}(\xi^{2})\end{eqnarray}
 This gives, once more, the divergence \begin{equation}
t(\sigma)\propto(\sigma_{c}-\sigma)^{-1/2}.\end{equation}

For a general threshold distribution the same procedure may be followed.
To use the exponential approach to the fixed point, as we have done,
may seem to be doubtful. But the rational is that for small $\sigma$
the starting point $n_{0}=1$ is already rather close to the fixed
point, while for larger $\sigma$ it doesn't matter much if the first
few iterations is not described well by the exponential formula, since
in any case the majority of the iterations occur close to the fixed
point.

\section{Concluding remarks}

A detailed numerical and analytic study of the relaxation dynamics
in finite fiber bundles subjected to external loads is presented.
The relaxation takes place in a number, $t(\sigma)$, of steps: In
each step all fibers weaker than the load per surviving fiber burst,
and the relaxation proceeds until equilibrium is reached or until
all fibers have failed. As function of the initial stress $\sigma$
the number of steps, $t(\sigma)$, shows a divergence $\left|\sigma-\sigma_{c}\right|^{-1/2}$
at the critical point, both on the subcritical and supercritical side.
This is a generic result, valid for a general probability distribution
of the individual fiber breakdown thresholds. 

On the supercritical side $t(\sigma)$ for large $N$, is independent
of the system size $N$. On the subcritical side there is, however,
a weak (logarithmic) $N$-dependence, as witnessed by eqs. ($46$),
($47$) and ($55$). 

The analytic estimates are based on the average strength of a group
of fibers. The comparison with the simulation data on the probabilistic
distribution of fiber strength shows that this is a satisfactory calculation
procedure. 

\vskip.2in

\begin{center}\noun{Acknowledgments}\par\end{center}

\vskip.1in

S. P. thanks the Research Council of Norway (NFR) for financial support
through Grant No. 166720/V30 and 177958/V30.

\end{document}